\documentclass[conference]{IEEEtran}
\IEEEoverridecommandlockouts

\usepackage{cite}
\usepackage{amsmath,amssymb,amsfonts}
\usepackage{algorithmic}
\usepackage{graphicx}
\usepackage{textcomp}
\usepackage{float}
\usepackage{multirow}
\usepackage{array}
\usepackage{multirow}
\usepackage[utf8]{inputenc}

\usepackage[utf8]{inputenc} 

\usepackage[hyphens]{url}            
\usepackage{booktabs}       
\usepackage{amsfonts}       
\usepackage{nicefrac}       
\usepackage{microtype}      
\usepackage{lipsum}		
\usepackage[numbers]{natbib}
\usepackage{doi}
\usepackage{amsmath} 
\usepackage{subcaption}
\usepackage[table,xcdraw]{xcolor}
\usepackage{array}
\usepackage{physics}
\usepackage{makecell}

\usepackage{cite}
\usepackage{hyperref}
\usepackage{url}

\graphicspath{{Copy_Thesis/}}

\def\BibTeX{{\rm B\kern-.05em{\sc i\kern-.025em b}\kern-.08em
    T\kern-.1667em\lower.7ex\hbox{E}\kern-.125emX}}
\begin{document}

\title{Introducing the Quantum Economic Advantage Online Calculator\\
\thanks{E.C.~was supported in part by ARO MURI (award No.~SCON-00005095), and DoE (BNL contract No.~433702). FutureTech funding provided by Accenture.} 
}

\author{\IEEEauthorblockN{1\textsuperscript{st} Frederick Mejia}
\IEEEauthorblockA{\textit{FutureTech Lab} \\
\textit{MIT}\\
Cambridge, MA, USA \\
fmejia@mit.edu}
\and
\IEEEauthorblockN{2\textsuperscript{nd} Hans Gundlach}
\IEEEauthorblockA{\textit{FutureTech Lab} \\
\textit{MIT}\\
Cambridge, MA, USA \\
hansgund@mit.edu, \href{https://orcid.org/0000-0001-5499-5072}{ORCID}
}
\and
\IEEEauthorblockN{3\textsuperscript{rd} Jayson Lynch}
\IEEEauthorblockA{\textit{FutureTech Lab} \\
\textit{MIT}\\
Cambridge, MA, USA \\
jaysonl@mit.edu}
\and
\IEEEauthorblockN{4\textsuperscript{th} Carl Dukatz}
\IEEEauthorblockA{\textit{Accenture Innovation} \\
\textit{Accenture}\\
Detroit, MI, USA \\
carl.m.dukatz@accenture.com}
\and
\IEEEauthorblockN{5\textsuperscript{th} Andrew Lucas}
\IEEEauthorblockA{\textit{FutureTech Lab} \\
\textit{MIT}\\
Cambridge, MA, USA \\
email address or ORCID}
\and
\IEEEauthorblockN{6\textsuperscript{th}Eleanor Crane}
\IEEEauthorblockA{\textit{MechE, MIT} \\
\textit{Belfer Center, Harvard Kennedy School}\\
Cambridge, MA, USA \\
emc2@mit.edu, \href{https://orcid.org/0000-0002-2752-6462}{ORCID}}
\and
\IEEEauthorblockN{7\textsuperscript{th} Prashant Shukla}
\IEEEauthorblockA{\textit{Accenture Research} \\
\textit{Accenture}\\
Boston, MA, USA \\
prashant.p.shukla@accenture.com}
\and
\IEEEauthorblockN{8\textsuperscript{th} Neil Thompson}
\IEEEauthorblockA{\textit{FutureTech Lab} \\
\textit{MIT}\\
Cambridge, MA, USA \\
neil\_t@mit.edu}
}
\maketitle

\begin{abstract}

Developing a systematic view of where quantum computers will outperform classical ones is important for researchers, policy makers and business leaders. But developing such a view is challenging because quantum advantage analyses depend not only on algorithm properties, but also on a host of technical characteristics (error correction, gate speeds, etc.). Because various analyses make different assumptions about these technical characteristics, it can be challenging to make comparisons across them. In this paper, we introduce an open-access web-tool designed to make such comparisons easy. Built on the framework introduced by Choi, Moses, and Thompson (2023), it calculates when quantum systems will outperform classical computers for a given algorithmic problem. These estimates can be easily updated based on various assumptions for error correction, overhead, and connectivity. Different hardware roadmaps can also be used and algorithm running times can be customized to particular applications. It can currently be accessed at \url{https://futuretech.mit.edu/quantum-economic-advantage-calculator}.

This integrated prediction tool also allows us to explore which technical factors are most important for quantum ``economic" advantage (outperforming on a cost-equivalent basis). Overall, we find that for some algorithms (e.g. Shor's) the timing of advantage is quite robust, whereas for others (e.g. Grover's) it is contingent, with numerous technical characteristics substantially impacting these dates. In the paper, we discuss both why this occurs and what we can learn from it.

\end{abstract}

\section{Introduction}

\label{sec:intro}

Jack Kilby, the inventor of Integrated Circuits, in his Nobel lecture outlined some of the earliest challenges in building electronic systems \cite{KilbyNobel}. He noted “the invention of vacuum tubes launched the electronics industry.” These electron tubes were the first hardware we used to manipulate the flow of electrons to our need. However, the use of electron tubes grew from radio broadcasting and being used as a switch to more complex applications that required integration of multiple parts. Then factors such as cost, space, switching speed, heat dissipation became the focus of electronic engineers working on integrated circuits\cite{moore1998cramming}. 

We find ourselves in a similar phase transition within quantum computing. Both science and market competition seem to be at work to determine the best hardware. Just as we moved from vacuum tubes, to Germanium, to eventually settling on silicon, a dominant design is perhaps in the offing. However, a slew of other parameters are also key determinants of feasibility and quantum economic advantage for specific business applications\cite{choi2023quantum,CMRQuantum}. This raises the question, which factors this time around, are most important? 

There is emerging literature looking at quantum advantage in the face of the complexity of quantum hardware and algorithms. Our analysis is based around the quantum economic advantage framework \cite{choi2023quantum}. Babbush et al \cite{babbush2021focus} identified the issue of finding quantum advantage with large hardware overheads. There has also been significant work trying to capture the limitation of quantum hardware to predict the date when critical security protocols like RSA-2048 will be broken with dates ranging from the late 2030s to mid-2040s \cite{sevilla2020forecasting}\cite{parker2023quantum_encryption}.

For instance, quantum computation holds immense potential to revolutionize fields such as cryptography, machine learning, molecular simulations, and more \cite{bobier2024long_term_forecast_quantum}. However, the journey from theoretical promise to practical utility is hindered by hardware limitations, including error rates, gate fidelity, and the slow performance of quantum systems compared to classical computers \cite{bobier2024long_term_forecast_quantum}. While the industry is investing heavily in quantum technology, the timeline for achieving Quantum Economic Advantage (QEA)—the point at which quantum systems become cost-efficient for solving specific problems—remains uncertain.

One of the key challenges in the quantum computing landscape is the diversity of qubit modalities, which complicates direct comparisons and assessments of quantum hardware capabilities. Quantum computers can be built using atoms, electrons, or photons, each with distinct subtypes offering unique advantages and challenges. For instance, trapped ions excel in coherence time but have slower gate speeds, while superconducting qubits offer faster gate speeds but have fidelity and scalability issues \cite{gschwendtner2023potential_challenges_quantum}. These nuances are often overlooked in mainstream discussions, making it difficult for non-experts to grasp the current capabilities of quantum systems.

Efforts to simplify comparisons, such as the quantum volume benchmark \cite{OliviaDiMatteo2020} or algorithmic runtime tests, provide some insight into live machine performance but fail to capture the future performance, scalability, and economic viability of quantum systems. The complexity of quantum hardware and the interplay of factors like qubit count, coherence time, gate speed, and error-correcting codes require a more comprehensive approach to evaluate when quantum systems can outperform classical computers.

To address this, the Quantum Economic Advantage Calculator (QEA Calculator) offers a powerful tool for predicting and estimating the viability of quantum computers. It allows users to input a wide range of parameters, including hardware specifications, algorithmic runtimes, error-correcting code ratios, physical qubit growth, etc to model the potential of quantum computing for specific problems. For non-experts, the tool provides default settings that simplify the process, offering increased clarity as to when quantum systems could become cost-efficient. Advanced users can fine-tune these parameters to conduct detailed comparisons of custom algorithms and runtime scenarios.

By integrating these diverse factors into a cohesive model, the QEA Calculator bridges the gap between the theoretical potential of quantum computing and its practical realization. It provides a framework for the quantum community to navigate the complexities of quantum technology. This tool not only fosters realistic expectations but also empowers researchers to allocate resources effectively, prioritize technologies with the potential for impact, and mitigate disillusionment within the industry. As quantum computing continues to evolve, tools like the QEA Calculator will play a crucial role in guiding the community toward practical and scalable applications.

\section{Quantum Advantage Model}
\label{sec:QEAModel}

\subsection{Foundations of the Quantum Tortoise and Classical Hare Framework}
\label{chap:original}

The tool we are introducing in this paper is an implementation and extension of the framework explicated in the Quantum Tortoise and Classical Hare paper by Choi, Moses, and Thompson \cite{choi2023quantum}. Within the framework, the authors define the key term “Quantum Economic Advantage” (QEA) for a particular problem as when a problem can be solved faster on a quantum computer than it could be on a comparatively expensive classical computer. 

To achieve quantum economic advantage, the QEA framework outlines two important conditions which must be reached. The first is feasibility. For it to be advantageous to switch to quantum hardware, the problem must first actually be executable on the hardware for a given problem size. This is primarily dependent on the number of qubits on the machine that algorithms can reliably run error-free. The second condition is algorithmic advantage, which is defined as the time when a quantum device could compute the solution to a problem (at a specific problem size) faster than an equally-priced classical machine. This condition relies on several factors, such as algorithmic and hardware speed differences. 

For a specific problem and quantum hardware, the feasibility and minimum problem sizes required for algorithmic advantage can be plotted over time. The space below the feasibility line represents the range of problem sizes which can practically be computed on hardware at that given year. The space above the advantage line represents the range of problem sizes for which it would be faster to perform those problems on the given quantum hardware. The point where the two lines overlap is called the quantum economic advantage point or QEA point. This marks the first problem size that is both feasible and advantageous to do on a quantum computer. We will also discuss the year of quantum economic advantage for some fixed problem size. In this case we want the first year for which solving the problem is feasible on a quantum computer and solving the problem of that size is faster on a quantum computer than classical computers. 

A basic view of our framework is based on five key inputs. These inputs are constants or functions of the problem size $n$ or time $t$, as follows:
\begin{enumerate}
    \item The roadmaps for the number of qubits in the quantum computer ($R(t)$)
    \item The ratio of physical to logical qubits needed for error correction, referred to as `Physical-Logical Qubit Ratio' (PLQR).
    \item The hardware slowdown between comparably expensive classical and quantum devices ($10^{\text{hws}}$), 
     referred to as 'Classical Compute Cost Advantage'.
    \item The runtime of the classical algorithm ($C(n)$)
    \item The runtime of the quantum algorithm ($Q(n)$)
\end{enumerate}

The minimum quantum advantageous problem size (referred to as $n^*$) is the problem size $n$ such that the problem’s classical runtime $C(n)$ equals the quantum runtime $Q(n)$ times the hardware slowdown of the device $10^{\text{hws}}$. This is:
$$n^* = n \text{ }| \text{ } C(n) = 10^{\text{hws}} * Q(n)$$

Likewise, the year of quantum economic advantage (referred to as $t^*$) is the year that the maximum feasible problem size equals the minimum advantageous problem size in that year. The maximum feasible problem size in a year for a specific roadmap is found by extrapolating the number of physical qubits available in that year and dividing them by the physical-logical qubit ratio to yield the number of logical qubits that year. We then use the qubit problem size function QPS$(q)$ \ref{Qubit to Problem Size} to determine the maximum problem size feasible on a quantum computer.

\section{Quantum Economic Advantage Calculator}

The Quantum Economic Advantage Calculator is a tangible model users can fully interact with to visualize how different outlooks on quantum progression could affect overall timelines for QEA. The tool serves to generate two key visualizations, which depend on the parameters associated with the specific problem-hardware instance in question. 

\begin{figure*}[ht]
\centering
\includegraphics[width=0.7\textwidth]{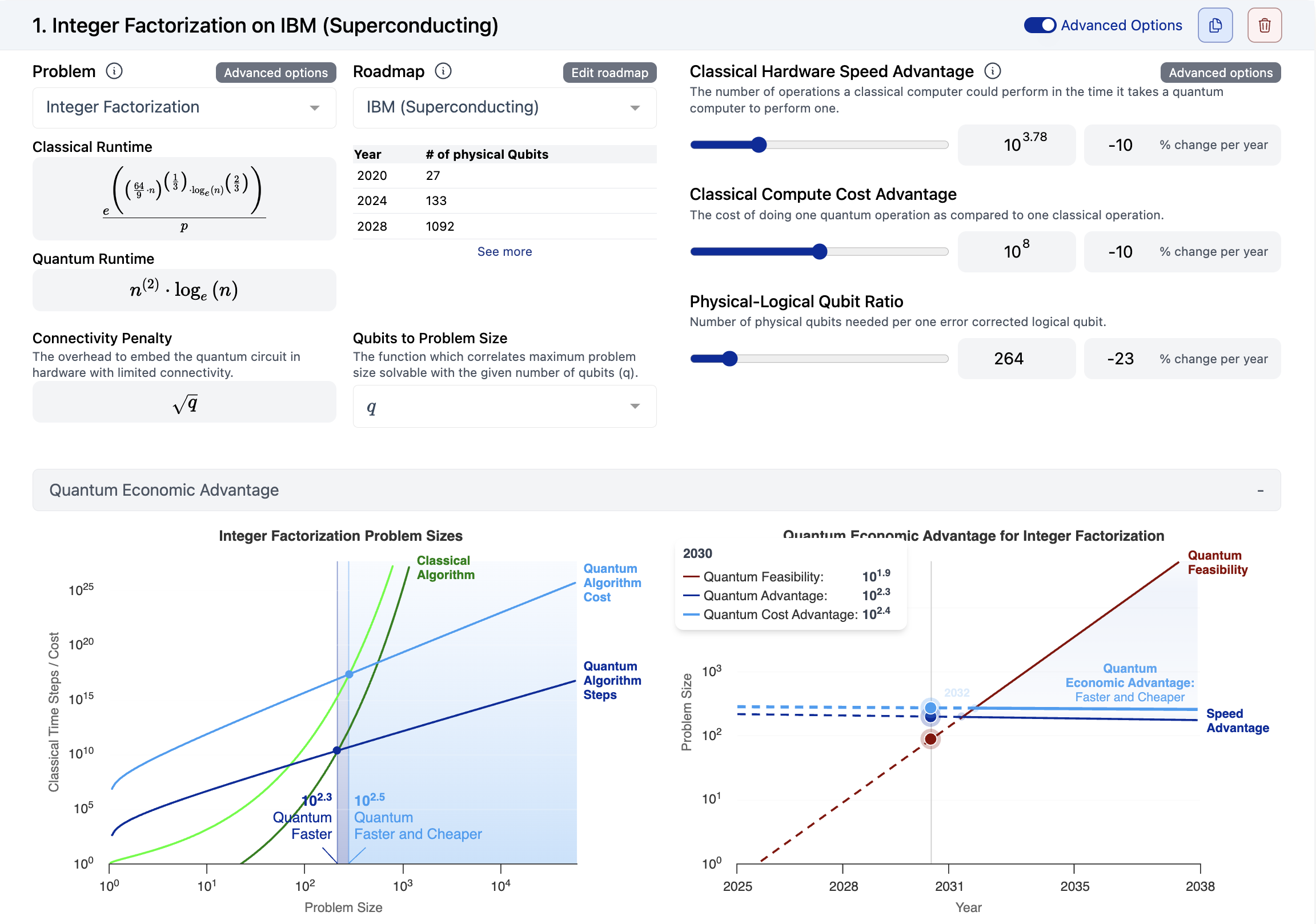}
\caption{User interface of the Quantum Economic Advantage Calculator. Users can adjust many parameters to fit their assumptions for quantum progress. The tool lets users view the quantum economic advantage point as well as the problems that have an advantage on a quantum computer. 
}
\label{fig:user_interface}
\end{figure*}

The first graph (bottom left of Figure \ref{fig:user_interface}) is of the minimum problem size needed for quantum algorithmic advantage ($n^*$). It contains trajectories of both the classical and quantum runtimes plotted with problem size on the x-axis and the amount of classical time steps\footnote{Since hardware constraints influence the relative efficiency of quantum operations compared to classical ones, the quantum runtime is normalized to the classical runtime when both are plotted on the same graph.} on the y-axis. The problem size where both lines intersect is $n^*$. 

The second graph in Figure~\ref{fig:user_interface} contains the most important information in the framework: when will the problem become economically advantageous to run on quantum hardware? This graph contains the plots of the feasibility and advantage lines discussed in the previous section. Their intersection marks the year where QEA occurs ($t^*$) on that hardware.

\subsection{Inputs to the Calculator}
\label{sec:inputs}

A summary of the input parameters can be found in Table~\ref{tab:parameters}. We will now discuss them in detail.

\subsubsection{Problem Inputs}

\paragraph*{Runtimes}
A primary input to the framework is the algorithmic problem being analyzed, as it determines the classical and quantum runtimes complexities under comparison. The tool provides preset options for users to explore, such as integer factorization, unstructured database search, and the traveling salesman problem, each incorporating the best-known worst-case time complexities for their respective computations.

The calculator also offers users complete flexibility to define and modify the runtime expressions. Users can adjust existing runtimes or input entirely new expressions to suit their analyses. For instance, an individual with an optimistic view of quantum algorithm development could input quantum runtimes that are significantly faster than the current state-of-the-art. 

\paragraph*{Parallelizability}
\label{sec:parallel}

An enhancement to the model, previously unaddressed, is the inclusion of parallelizability. For some problems, classical computers can distribute their workload across multiple processors, reducing overall runtime~\cite{ravi4742023parallel}. Overlooking, this capability would underestimate the performance of classical devices and affect the accuracy of comparisons between classical and quantum systems.
This is one area where the `cost comparable' part of quantum economic advantage can be most clearly included.

To address this, our framework allows classical runtimes to be expressed as functions of $n$ (problem size) and $10^p$ (number of available processors). Users can define runtimes as such:
$$C(n, p) = \frac{C_{\text{parallelizable}}(n)}{10^p} + C_{\text{nonparallelizable}}(n)$$

Although the runtime decreases with the number of processors, the total number of computations does not decrease. We call the number of operations performed as the "classical work." It is equivalent to the expression for runtime when evaluated with a single processor, $C(n,1)$.

Parallelization in terms of multiple processors for quantum algorithms is not considered in our framework. The challenges of maintaining entanglement and coherence make parallelization in quantum devices significantly more complex. While there is ongoing research aimed at enabling quantum parallelism~\cite{parekh2021quantum}, practical solutions are not yet feasible, so we have excluded it from our analysis. However, we do allow users to input both a quantum runtime and work; so adjustments to the quantum runtime to reflect the level of parallelism is possible.

\paragraph*{Qubit to Problem Size}\label{Qubit to Problem Size}
An important part of our framework is  the Qubit to Problem Size (QPS) function . 
We refer to this function as Qubit to Problem Size (QPS). Different quantum problem sizes have different qubit requirements in relation to the problem size solved. For example, for Shor's factorization algorithm, the number of qubits scales linearly with the number of digits in the factored number \cite{beckman1996efficient}. While for Grover's search algorithm, the number of qubits necessary is only the log of the search space. 

To account for these differences, the calculator provides three selectable QPS options: exponential ($2^q$), linear ($q$), and logarithmic ($\log_2(q)$). While these options are not as customizable as the runtime expressions, they collectively span the key hierarchies of how problem size relates to the number of logical qubits. 

\subsubsection{Hardware Inputs}
\paragraph*{Roadmap}
\begin{figure*}[ht]
\centering
\includegraphics[width=0.7\textwidth]{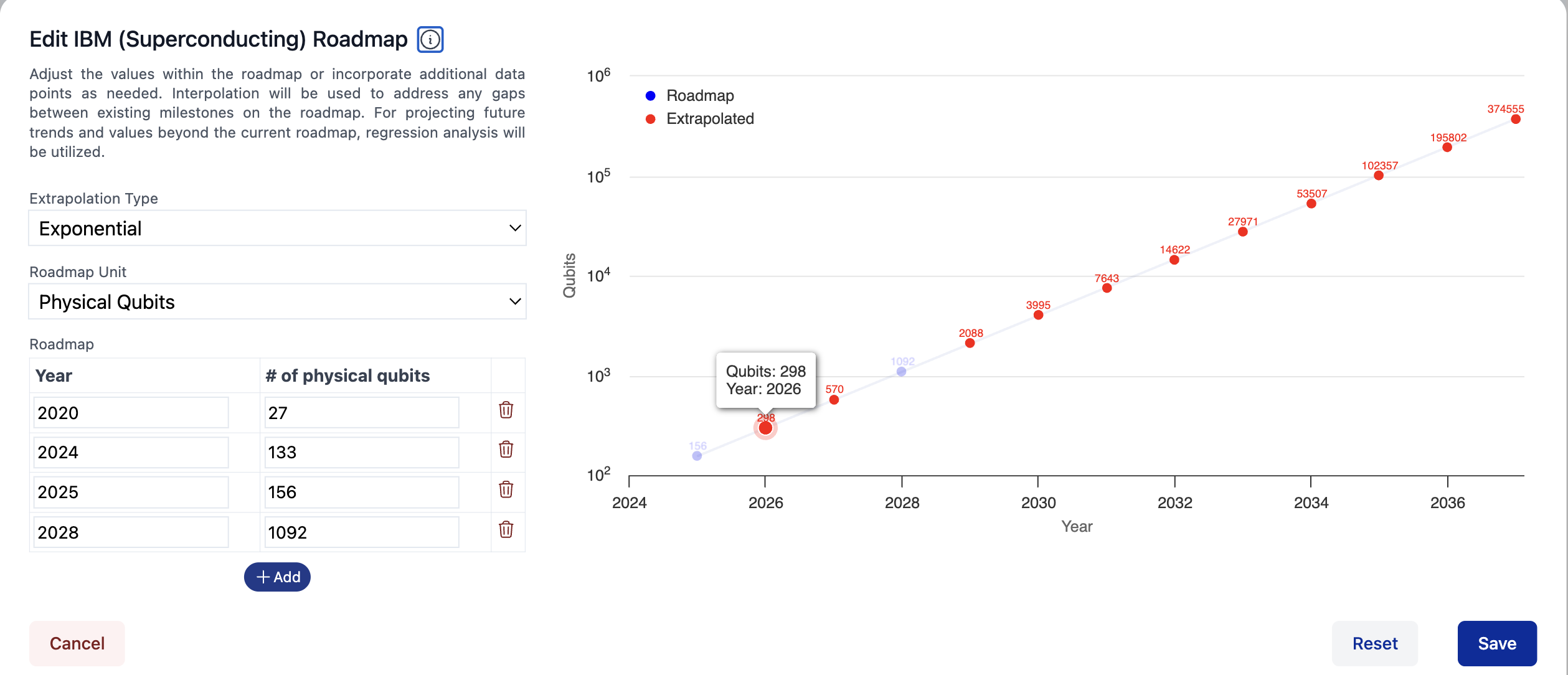}
\caption{Our QEA tool comes with preregistered roadmaps. User also have the ability to modify or customize their own quantum roadmaps. Blue dots are from the roadmap and red dots are an exponential extrapolation based on that roadmap.
}
\label{fig:editing_roadmap}
\end{figure*}

The primary hardware inputs are the quantum roadmaps used for qubit extrapolation.

The calculator has preset options for quantum hardware providers such as IBM, Google, IonQ, and Pasqal. Each roadmap outlines projections for physical qubits along with target years, based directly on the providers' published timelines \cite{ibm_quantum_roadmap,google_quantum_roadmap,chapman_scaling,pasqal_roadmap_2024}.
The calculator allows users to modify the existing roadmaps directly. Qubit-year pairs can be adjusted to align with different expectations, and entirely new points can be added to the projection. If an optimistic individual is confident that quantum devices will exceed a certain amount of physical qubits by a certain year, they have the freedom to update points in the roadmap. All calculations will automatically update to reflect the faster growth. Alternatively, if a corporate roadmap looks overly ambitious one can correct this.

By default, the tool estimates exponential growth in qubits for each roadmap. When predicting the number of available qubits, the tool interpolates values between existing points exponentially and extrapolates beyond the final data points using the same growth rate. Users can also choose linear growth instead, in which case the tool interpolates and extrapolates linearly.
These projections for physical qubits are ultimately used to estimate the number of logical qubits, which then feed into further calculations. Alternatively, users can bypass this step by directly inputting a roadmap for logical qubits. In this case, the tool extrapolates the number of logical qubits directly from the user-defined points, skipping the intermediary calculations involving physical-logical qubit ratios.

\paragraph*{Hardware Slowdown}
Hardware slowdown, as previously described, refers to the number of operations a classical computer can perform in the time it takes a comparably priced quantum computer to perform a single operation. Since this factor can vary across multiple orders of magnitude, users input this value logarithmically, as a power of 10.

The hardware slowdown parameter can be directly input into the calculator. However, the framework also allows users to compute this value by specifying the individual factors that contribute to it. In the original framework, hardware slowdown was expressed as the product of three factors:
\begin{enumerate}
    \item Speed Ratio: The ratio of the speed of a classical computer to that of the quantum computer.
    \item Gate Overhead: A factor representing the additional gates (operations) required to maintain fault tolerance.
    \item Algorithmic Constant Ratio: The ratio of the multiplicative constant from the classical algorithm's runtime to that of the quantum algorithm's.  
\end{enumerate}

Like the hardware slowdown, the speed ratio can either be entered directly into the calculator or calculated from individual factors. Users can provide the 2-qubit gate time of the quantum computer and the clock speed of the classical device, which is assumed to be 5 GHz by default\footnote{While 5 GHz is typical for consumer machines~\cite{IntelClockSpeed}, it is optimistic for classical supercomputers, which prioritize parallelism over clock speed~\cite{nasa_pleiades}. However, as will be discussed in the later analysis, this assumption has minimal impact on the model's output.}. The tool will then compute the speed ratio based on these specifications.

\begin{figure}[ht]
\centering
\includegraphics[width=0.4\textwidth]{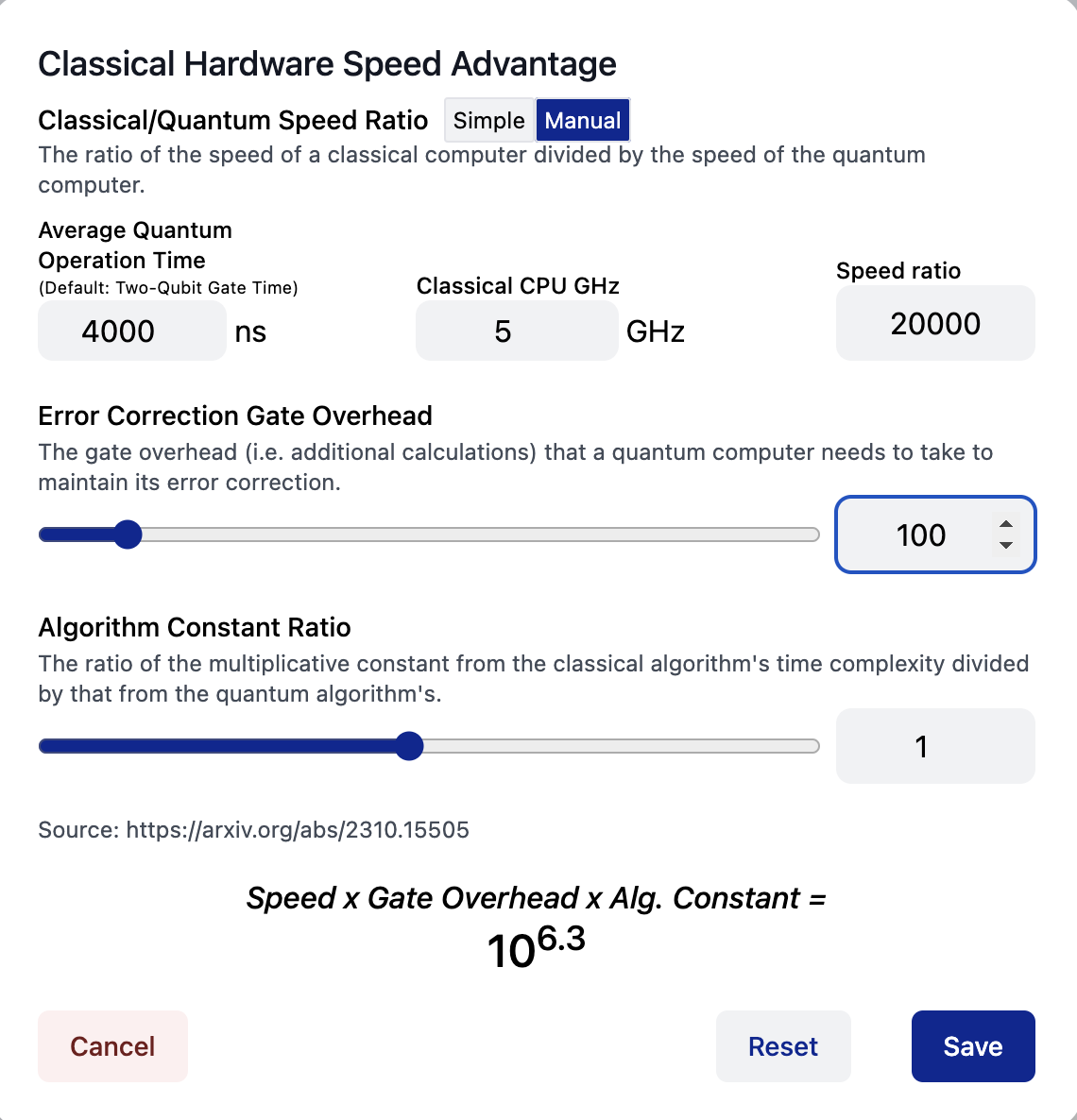}
\caption{Users can adjust hardware and algorithmic factors to determine the quantum speed advantage. 
}
\label{fig:your_label}
\end{figure}

\paragraph*{Quantum Improvement Rate}
The quantum improvement rate (QIR) represents this framework’s method of modeling changes in the hardware slowdown over time, expressed as the percentage improvement year-to-year\footnote{In the calculator, improvements are associated with negative values (decreasing overhead), while increases are represented with positive values.}. With the recent rapid advancements in quantum hardware and the slowing of Moore’s law’s exponential growth in classical computation~\cite{preskill2012quantum}, it is reasonable to assume that the hardware slowdown factor will decrease over time.

QIR is used to calculate the annual reduction in the hardware slowdown. Representing the hardware slowdown as $S_t = 10^{\text{hws}_t}$, its change over time can be modeled by\footnote{The decay of hardware slowdown is capped to prevent the model from assuming that quantum machines will eventually outperform comparably priced classical machines in terms of speed.}:
$$S_{t+1} = S_t * (1 + \text{QIR}/100)$$
It is worth noting that users are also free to input a positive change per year, implying that quantum devices are becoming relatively slower over time.  

\paragraph*{Physical-Logical Qubit Ratio}
The physical-logical qubit ratio (PLQR) represents how many physical qubits are needed to encode a single logical qubit. Its value can be directly input by the user into the calculator.

\paragraph*{Physical-Logical Qubit Ratio Improvement Rate}
Just as the hardware slowdown evolves based on the quantum improvement rate, the PLQR is influenced by the physical-logical qubit ratio improvement rate (RIR\footnote{Represented mathematically as RIR instead of PLQRIR for clarity and brevity.}). This adjustment is input and calculated in the same manner as the quantum improvement rate\footnote{The decay of PLQR is limited to a minimum value of 3, reflecting the constraints imposed by the most basic error correction schemes~\cite{cambridge_quantum_error_correction}.}.

\paragraph*{Connectivity Penalty}
\label{sec:connectivity}
Connectivity refers to the arrangement of interactions between qubits within a quantum system, often represented as a qubit interaction graph, where vertices correspond to qubits and edges denote allowable direct interactions. In practice, not all qubits in a quantum device can interact directly, and this limited connectivity imposes constraints on quantum algorithm design~\cite{holmes2018impact}.

How connectivity is handled largely depends on the underlying quantum architecture. For example, in superconducting machines (such as the IBM Quantum Heron~\cite{ibm_marrakesh}), qubits are typically arranged in 2D grids, whereas trapped ion qubits are typically arranged in linear chains~\cite{ionq_technology}.

The level of connectivity in a quantum device directly influences the efficiency of implementing quantum algorithms on the hardware. When an algorithm requires interactions between qubits that are not directly connected, additional operations are necessary to route information, which increases computational overhead.

In our framework, connectivity overhead is modeled as a multiplicative penalty on the quantum runtime. For hardware with limited connectivity, we apply a default penalty of $\sqrt{q}$, where $q$ represents the number of logical qubits required for the computation. This aligns with the asymptotic overhead if many operations are nonlocal which scales as $O(\sqrt{q})$\footnote{This depth overhead assumes connections are bound to 2D space.}~\cite{herbert2018depth}.

Since the relationship between circuit connectivity and runtime is complex, we designed this parameter to be as freely customizable as the runtimes themselves. The tool accepts expressions in terms of $q$, allowing users to adjust connectivity penalties to reflect the specific constraints of their quantum hardware.

\paragraph*{Processors}
This input represents the number of processors ($10^p$) that would exist in a classical machine that is comparably priced to the quantum one being considered. As elaborated in the parallelizability section (\ref{sec:parallel}), the number of processors reduces the time to perform a classical computation. Like the hardware slowdown, it is input logarithmically.

\paragraph*{Cost Improvement Rate}
Our framework incorporates an additional parameter for the cost improvement rate (CIR), which captures the relative rates of cost reduction for quantum and classical operations over time. Specifically, it is defined as the ratio of the annual rates of change in the costs of quantum and classical operations. For instance, if quantum costs decrease at a rate of 30\% per year (0.7 multiplier), while classical costs decrease at a rate of 20\% per year (0.8 multiplier), the ratio of change would be 7/8. This implies that quantum costs improve at an effective rate of approximately 12.5\% per year relative to classical costs.

While performance improvements influence the hardware slowdown associated with quantum runtimes in the model, cost reductions primarily determine the number of classical processors that can be leveraged to speed up classical runtimes. If quantum costs decline more rapidly than classical costs, the cost-equivalent classical system becomes smaller over time, as fewer classical processors can be purchased for the same budget. Consequently, in the tool, the number of classical processors dividing into the classical runtime decreases each year. Both cost and performance improvements ultimately impact the problem size required for algorithmic advantage. This assumption applies when quantum costs are improving faster than classical costs, though users have the flexibility to modify or disregard it if needed. 

\subsection{Calculations}
\label{sec:math}
With all of the calculator’s inputs introduced, we can finally discuss how the tool calculates its findings. Before we dive into the mathematics, let us present Table \ref{tab:parameters} showing how each input is expressed in the equations. 

\begin{table*}[ht]
\centering
\resizebox{\textwidth}{!}{
\begin{tabular}{|c|c|p{0.5\textwidth}|}
\hline
\textbf{Parameter} & \textbf{Symbolic Representation} & \textbf{Context} \\
\hline
Problem Size & $n$ & The parameter used to calculate runtimes and necessary logical qubits. \\
\hline
Number of Processors & $10^p$ & A factor dividing into the parallelizable classical runtime. \\
\hline
Classical Runtime & $C(n, 10^p)$ & A function in terms of problem size and the number of processors.\\
\hline
Quantum Runtime & $Q(n)$ & A function in terms of problem size.\\
\hline
Qubit to Problem Size & QPS & Function used to calculate the maximum feasible problem size from the number of logical qubits. Its inverse QPS$^{-1}$ is often used.\\
\hline
Roadmap & R(t) & Function which outputs the expected number of physical qubits at the given year, considering the means of extrapolation and the growth of the roadmap.\\
\hline
Hardware Slowdown & $10^{\text{hws}}$ & The relative slowdown on quantum computers.\\
\hline
Quantum Improvement Rate & QIR & The rate at which the hardware slowdown decreases.\\
\hline
Physical-Logical Qubit Ratio & PLQR & The number of physical qubits required to represent one logical qubit.\\
\hline
Physical-Logical Qubit Ratio Improvement Rate & RIR & The rate at which the physical-logical qubit ratio decreases.\\
\hline
Connectivity Penalty & $P(q)$\footnotemark & A function in terms of the number of logical qubits.\\
\hline
Cost Improvement Rate & CIR & The rate at which the number of processors decreases.\\
\hline
\end{tabular}
}
\caption{Framework Parameters and Their Symbolic Representations}
\label{tab:parameters}
\end{table*}
\footnotetext{The number of logical qubits $q$ is expressed as QPS$^{-1}(n)$.}

As a reminder, the minimum problem size for algorithmic advantage is the smallest problem size where the runtime on quantum hardware is faster than the runtime on classical hardware. 
The final expression for determining $n^*$ (the minimum problem size for advantage) is based on the hardware slowdown, parallel runtime of the classical algorithm, the quantum runtime, the number of available processors, and the connectivity penalty. It is given by:
$$n^* = n \text{ }|\text{ } C(n, 10^p) = 10^{\text{hws}} * Q(n) * P(\text{QPS}^{-1}(n))$$

Note that this expression calculates the \emph{current} problem size needed for advantage. The framework also incorporates the improvement of variables over time, with the trend of the required $n^*$ to maintain advantage forming the Advantage Line in the QEA Graph.
With the inclusion of the variables which account for rates of change, the generic expression for the advantage line given a year $t$ (which we’ll label Adv($t$)) becomes:
\begin{align*}
\text{Adv}(t) &= n \text{ }|\text{ } C(n, 10^p * (1 + \text{CIR} / 100) ^{t - t_0}) \\
&= 10^{\text{hws}} * (1 + \text{QIR} / 100) ^ {t - t_0} * Q(n) * P(\text{QPS}^{-1}(n))   
\end{align*}
where $t_0$ is the current year.
The calculator must also model feasibility over time. Given a user’s chosen roadmap, which includes the number of qubits achieved at target years and the extrapolation method, the tool calculates the maximum feasible problem size. It also considers how many physical qubits are needed for logical qubits, their expected rate of change, and the conversion from logical qubits to the maximum solvable problem size. Using these factors, the tool computes the maximum feasible problem size at a given year $t$ (Feas($t$))\footnote{PLQR and its improvement are disregarded if the roadmap provided was already in terms of logical qubits.}:
$$\text{Feas}(t) = \text{QPS}\left(\frac{R(t)}{\text{PLQR} * (1 + \text{RIR} / 100)^{t - t_0}}\right) $$ 
With these formulations for how the problem size needed for advantage and the maximum feasible problem size change over time, we can express the time of Quantum Economic Advantage as their intersection. That is:
$$t^* = t \text{ }|\text{ } \text{Feas}(t) = \text{Adv}(t)$$
If no advantage is found by the year 3000 with the given parameters, the tool will inform the user that no QEA exists. Conversely, if the $n^*$ required for advantage is already less than the current feasible problem size, the tool will indicate that QEA has already been achieved.

\subsection{Cost Advantage}
\label{seb:Cost Advantage}
All of the insights provided by the calculator so far have been focused on determining when a specific instance of an algorithm would execute faster on equivalently priced quantum hardware compared to classical hardware. The calculator also provides the ability to determine when certain algorithms will become cheaper to execute on quantum hardware, even if this might take longer. This is achieved by adapting the previous expressions with new variables.

\subsubsection{Cost Inputs}

For speed comparisons, the tool evaluates quantum and classical hardware by comparing their runtimes, factoring in a hardware slowdown. For cost comparisons, the tool assesses the total computational effort required by each hardware, where the computational effort corresponds to the total number of operations required to complete the computation. These operations are directly related to the original problem's runtimes.

As previously mentioned, classical work ($C_w$) is defined as the number of operations associated with the classical runtime when evaluated using a single processor. By default, classical work is evaluated as $C(n, 1)$, but users have complete flexibility to change it to a different expression.

Quantum runtime reflects the depth of the circuit implementing the algorithm. To quantify the total computational effort, quantum work ($Q_w$) is defined as the product of the runtime and the number of logical qubits involved in the computation. Quantum work is expressed using an additional variable $q$, which represents the number of logical qubits. By default, quantum work is the product of the quantum runtime and $q$. The value of $q$ used in the expression is $\text{QPS}^{-1}(n)$, reflecting how the required number of qubits changes with problem size.

\begin{figure}[ht]
\centering
\includegraphics[width=0.4\textwidth]{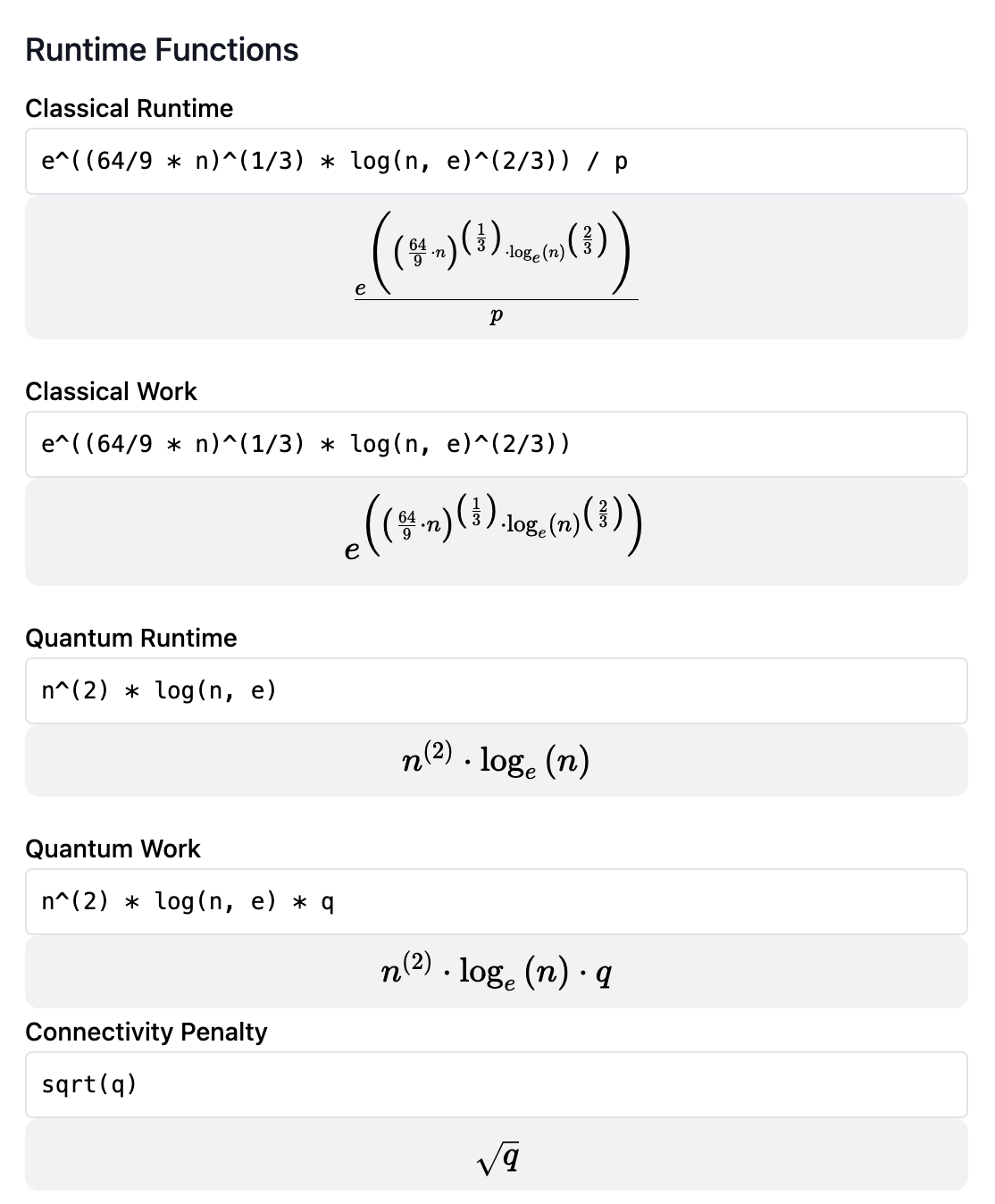}
\caption{Changing expressions for runtimes, work, and the connectivity penalty. 
}
\label{fig:runtimes}
\end{figure}

The final adjustment for cost comparison involves incorporating the cost factor ($10^{\text{cf}}$), which represents how much more expensive it is to perform operations on a quantum machine compared to a classical one. This cost factor is analogous to the hardware slowdown for speed and is input logarithmically. The cost factor can also vary over time, changing at a rate determined by the cost improvement rate, as discussed earlier.

\subsubsection{Cost Calculations}
The minimum problem size at which quantum computation becomes cheaper can be determined using the following expression:
$$n_c^* = n \text{ }|\text{ } C_w(n) = 10^{\text{cf}} * Q_w(n, \text{QPS}^{-1}(n)) * P(\text{QPS}^{-1}(n))$$

\noindent The trend in $n_c^*$ over time defines the cost advantage, which is expressed as Adv$_c(t)$:
$$\text{Adv}_c(t) = n \text{ }|\text{ } C_w(n) = 10^{\text{cf}} * (1 + \text{CIR} / 100)^{t - t_0} $$
$$* Q_w(n, \text{QPS}^{-1}(n)) * P(\text{QPS}^{-1}(n))$$
Since feasibility is not affected by cost considerations, the time at which it becomes cheaper to perform the problem on quantum hardware is determined by the intersection of Adv$_c(t)$ and Feas$(t)$:
$$t_c^* = t \text{ }|\text{ } \text{Feas}(t) = \text{Adv}_c(t)$$

\section{Results and Analysis}
\label{chap:analysis}

To better understand the practical usage of the QEA Calculator we performed comparative and sensitivity analysis on two different algorithmic problems on three different types of hardware. We chose integer factorization and search as particularly important and representative problems. 

\paragraph*{Integer Factorization}
Integer factorization is pivotal in quantum computing due to its role in cryptography. Shor’s algorithm for factoring numbers was groundbreaking because of its potential efficiency to break widely used RSA encryption. This has made it a hallmark problem for assessing quantum advantage and its potential implications for global data security. Its runtimes are:\\ 
Classical Runtime~\cite{PomeranceGNFS}: $e^{(\frac{64}{9}n)^\frac{1}{3} * (\ln n) ^ \frac{2}{3}}$\\
Quantum Runtime\footnote{A slightly faster runtime exists than was used in this analysis~\cite{regev2023}, but due to the small difference between the runtimes, the estimates for QEA would remain largely unaffected.}~\cite{shor1994algorithms}: $n^2 * \ln n$

\paragraph*{Database Search}
Searching through unstructured data is a widely relevant computational challenge \footnote{Unstructured search on classical data ie database search requires QRAM \cite{Giovannetti_2008} which may be impractical}. Typical search over an unordered set classically is worst case linear time with respect to the input.  Grover's algorithm highlights quantum computing's potential by providing a quadratic speedup over classical methods. The runtimes are:\\
Classical Runtime: $n$\\
Quantum Runtime~\cite{grover1996}: $\sqrt{n}$

\noindent In our analysis, we will assume that the classical runtimes for each problem are fully parallelizable, meaning that doubling the number of processors would halve the total runtime.

\subsection{Hardware Instances}
The quantum hardware selected for the analyses were chosen to cover different means of implementing quantum computers (superconducting, ion trap, and neutral atom). The three providers chosen were IBM (superconducting), IonQ (ion-trap), and QuEra (neutral atom).

\paragraph*{IBM}
IBM’s quantum computers are based on superconducting qubits which offer faster gate speeds compared to other architectures. Superconducting machines have exhibited 2-qubit gate times of 12 ns~\cite{martinis2019quantum}, which when converted using the formula from the inputs section, yields a hardware slowdown of approximately 6000 ($10^{3.78}$).  However, these faster gate times come at the cost of limited connectivity~\cite{ibm_marrakesh}, which we set as the default outlined in Section \ref{sec:connectivity}.

\paragraph*{IonQ}
IonQ’s quantum computers utilize trapped ion technology which features superior qubit connectivity. Their devices contain all-to-all connectivity meaning each qubit is able to directly interact with all others without the need for additional operations. As a result, IonQ will not have an associated connectivity penalty within the framework. However, ion trap quantum computers generally have much slower 2-qubit gate speed at around 600,000 ns~\cite{ionq_aria} (a slowdown of $10^{8.48} \approx 3.02 * 10^8$).

\paragraph*{QuEra}
The final quantum hardware provider, QuEra, leverages neutral atom technology, which uses arrays of ultracold atoms which can be dynamically rearranged to optimize qubit interactions. QuEra has a 2-qubit gate speed of 250 ns\footnote{This value is converted from its 4 MHz gate speed.}, which places it in the middle of the three hardware in terms of speed (yielding a slowdown of $10^{5.1} \approx 126,000$). The provider also features all-to-all connectivity meaning it also will not contain this runtime penalty~\cite{quera_onprem_2023}. 

\paragraph*{Assumptions and Parameter Estimations}
To provide a consistent basis for unknown parameters in the analyses, several assumptions and parameter estimations were made regarding the improvement rates, processor counts, and other hardware-specific values.

Since IBM’s roadmap lacks the physical number of qubits for the year 2029 (and only contains the logical number), its value was approximated by taking an extrapolation of the change in the physical-logical qubit ratio and using it on the known logical value. In 2023, the PLQR was around 400~\cite{abbasi2023understanding}, and by 2033, IBM's device is expected to have a PLQR of 50~\cite{ibm_quantum_roadmap}. Extrapolating these values indicated an improvement rate of about 23\% per year for the PLQR. Given that this rate is derived from real machines and other hardware platforms do not provide data to infer their yearly PLQR improvement, we will assume that all hardware will improve their PLQR at the same rate.

\begin{table}[ht]
\centering
\begin{tabular}{|p{3.5cm}|c|c|c|}
\hline
\textbf{Parameter} & \textbf{IBM} & \textbf{IonQ} & \textbf{QuEra} \\
\hline
Hardware Slowdown & 3.78 & 8.48 & 5.1 \\
\hline
Quantum Improvement Rate (quantum slowdown decline) & $10\%$ & $10\%$ & $10\%$ \\
\hline
Connectivity Penalty & $\sqrt{q}$ & 1 & 1 \\
\hline
Physical-Logical Qubit Ratio & 264 & 32 & 100 \\
\hline
\shortstack[l]{Physical-Logical Qubit \\ Ratio Improvement Rate \\ (fall in physical-to-logical ratio)} & $23\%$ & $23\%$ & $23\%$ \\
\hline
\shortstack[l]{Cost Improvement Rate\\ (decrease in cost overhead)} & $10\%$ & $10\%$ & $10\%$ \\
\hline
Processors & $10^8$ & $10^8$ & $10^8$ \\
\hline
\end{tabular}
\caption{Starting Parameters for IBM, IonQ, and QuEra Hardware in the Analyses. Note that in the online calculator, improvement parameters are negative to indicate a decline in quantum overhead.}
\end{table}

In the analyses, the cost improvement rate and quantum improvement rate were both set to a default value of 10\% (i.e 10\% decline in quantum slowdowns and cost overhead per year) for all hardware and the number of processors dividing the classical runtime was set to a value of $10^8$ \cite{chemistry_advantage}. These values were chosen as reasonable and illustrative starting points and enable a consistent baseline for exploring the framework's behavior and conducting sensitivity analyses.

\section{Key Findings}

\begin{figure}[ht]
\centering
\includegraphics[width=0.5\textwidth]{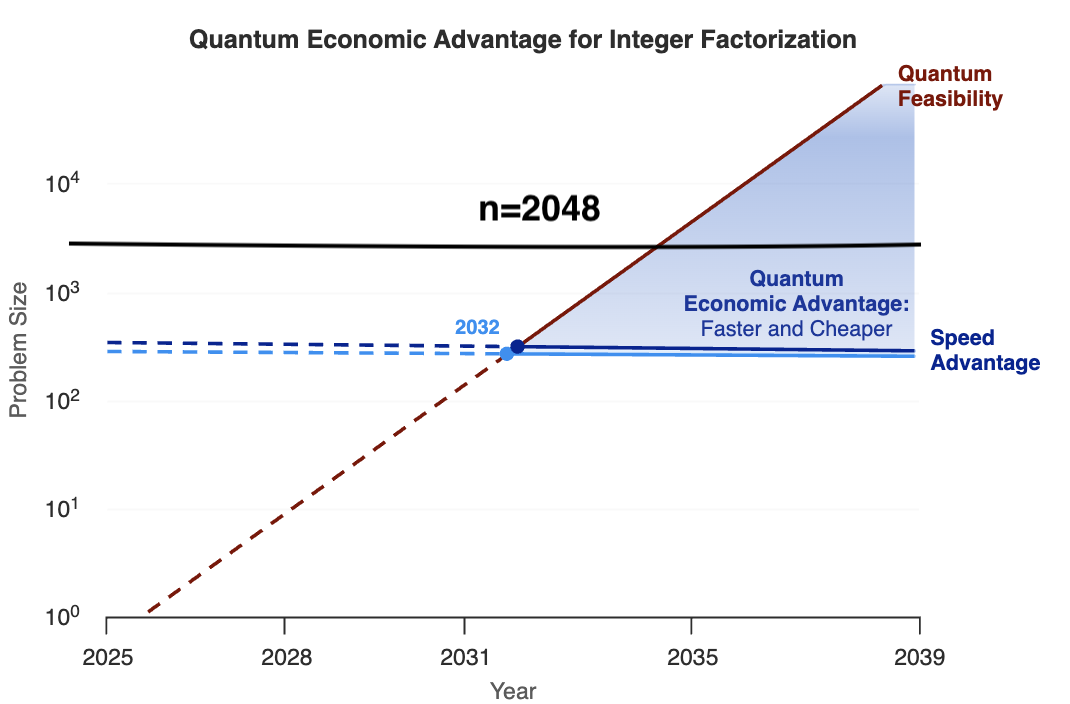}
\caption{Projected Quantum Economic Advantage for Integer Factorization on IBM Hardware (taken from online calculator). Extrapolating the maximum feasible problem size into the future (based on roadmap projections and assumptions for PLQR and RIR) implies that IBM hardware will be capable of performing factorization on problem sizes large enough to break 2048-bit encryption around 2034.}
\end{figure}

We present the quantum economic advantage years for the problems in our analysis. These represent the first year a quantum computer can solve some problem more efficiently than a classical computer. These do not represent the first year a quantum computer can solve a particularly useful problem size. We include this analysis in section \ref{Robustness Analysis}.

\begin{table}[ht]
\centering
\resizebox{0.5\textwidth}{!}{%
  \begin{tabular}{|l|c|c|c|}
    \hline
    \textbf{Company} & 
    \shortstack{\textbf{QEA Year} \\ \textbf{Search}} & \shortstack{\textbf{QEA Year} \\ \textbf{Factorization}} & \textbf{RSA‑2048 Year} \\ \hline
    IBM   & 2030 & 2032 & 2034 \\ \hline
    IonQ  & 2030 & 2029 & 2033 \\ \hline
    QuEra & 2025 & 2026 & 2028 \\ \hline
  \end{tabular}%
}
\caption{Predictions based on vendor roadmaps. Quantum error advantage (QEA) predictions for factorization and search tasks, alongside the year when RSA‑2048 can be cracked on that vendor roadmap. We view the vendor roadmaps, particularly for QuEra as overly optimistic.}
\label{tab:qea_rsa_predictions}
\end{table}

\subsection{Robustness Analysis}\label{Robustness Analysis}
Our robustness analysis looks at the effect of changes in our model's parameters on the years in which quantum advantage is achieved for relevant problem sizes. We focus on the two most relevant algorithms Shor's algorithm and Grover's Search. These two algorithms represent the character of our model with exponential and polynomial speedups, respectively. For Shor's algorithm, we focus on a problem size of 
\textbf{2048}. The year at which Shor's algorithm is capable of solving this problem size is of immediate security importance because it would result in cracking the internet security protocol RSA-2048. For Grovers algorithm we choose a problem size of $\mathbf{10^{20}}$. This number represents the rough upper bound of hashes a future state-of-the-art Bitcoin mining center with $10^{5}$ Bitcoin mining ASIC chips would be able to compute in 1 second \footnote{A current mining ASIC can compute roughly $10^{14}$ hashes a second \cite{Minerstat2025AntminerS19Pro}. We multiply by 10 to account for hardware progress. One mining-ASIC can have many hundreds of cores \cite{alcorn2022intel}. Therefore, $10^{5}$ ASICs is consistent with our $10^8$ parallelism overhead. }. If a quantum computer could search through hashes faster than this then it would be economical to replace Bitcoin mining rig centers with quantum computers. Note that an input size of $10^{20}$ is on the order of 100 Exabytes so problems impacted by this quantum algorithm must be exceptionally large.

\begin{figure*}
\centering
\begin{subfigure}{.45\textwidth}
  \centering
  \includegraphics[width=\linewidth]{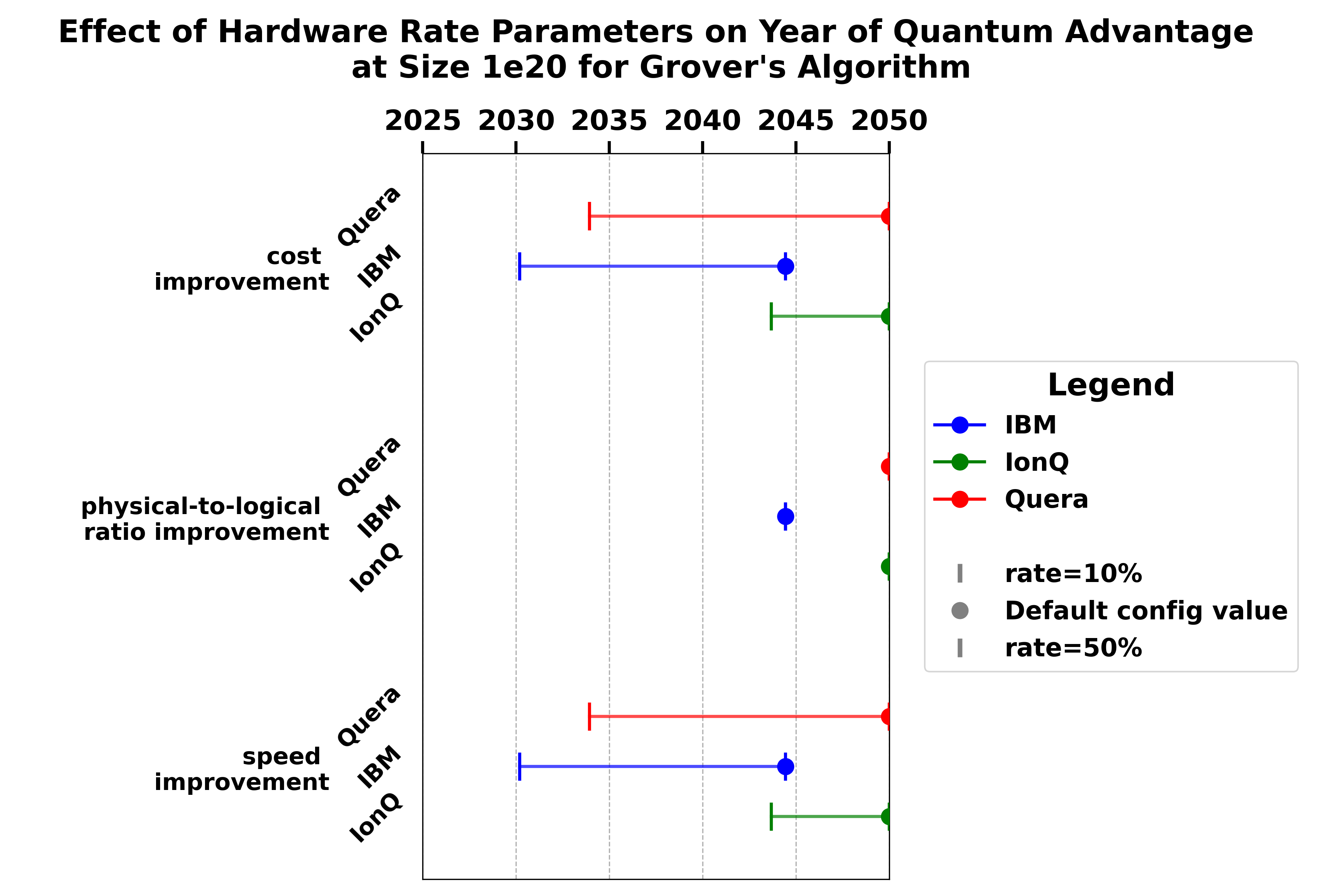}
\end{subfigure}%
\begin{subfigure}{.45\textwidth}
  \centering
  \includegraphics[width=\linewidth]{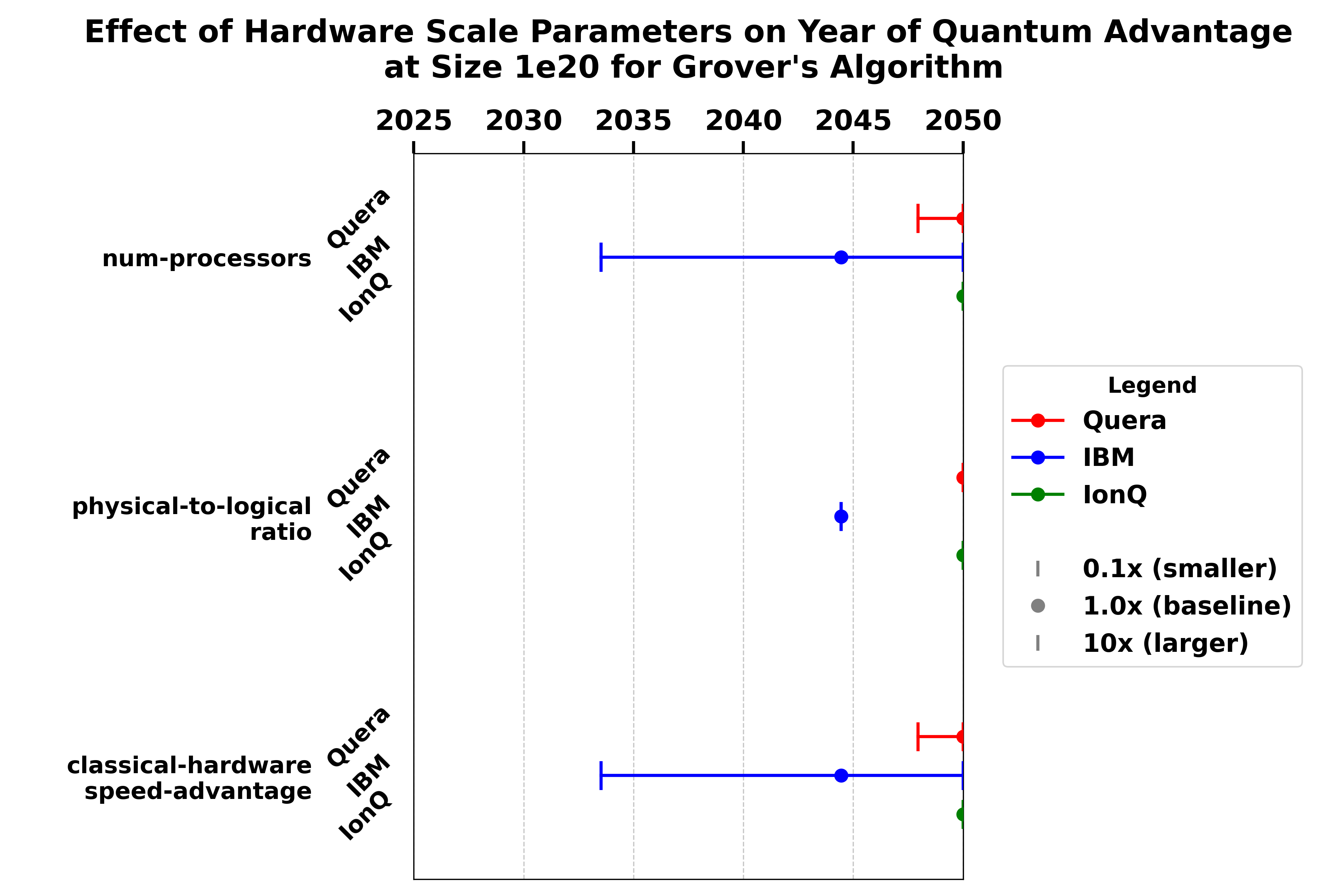}
\end{subfigure}
\caption{
Robustness studies for Grover's algorithm using vendor roadmaps. The plot shows the year at which Grover's algorithm has an advantage at a problem size of 1e20 (i.e is physically feasible and economically preferable) for various hardware roadmaps. Bars show how the predicted year changes when that parameter changes from the baseline. The chart on the left covers rates of improvements (as \% per year) at the baseline. The right chart shows the classical speed advantage, number of processors, and physical to logical qubit ratios changed by a factor of 10 smaller and larger.}
\label{fig:grover_parameter_plots}
\end{figure*}

\begin{figure*}
\centering
\begin{subfigure}{.45\textwidth}
  \centering
  \includegraphics[width=\linewidth]{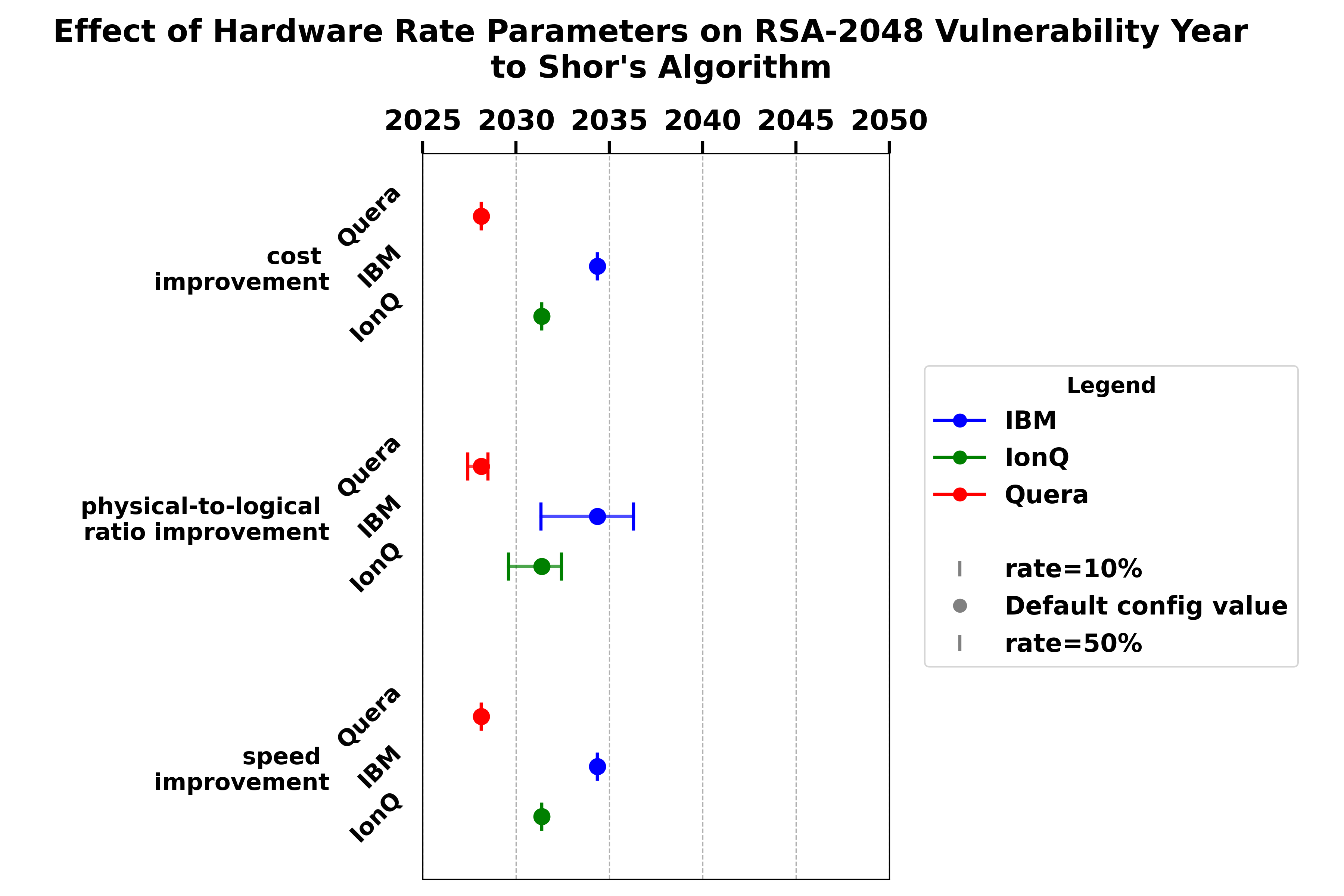}
\end{subfigure}%
\begin{subfigure}{.45\textwidth}
  \centering
  \includegraphics[width=\linewidth]{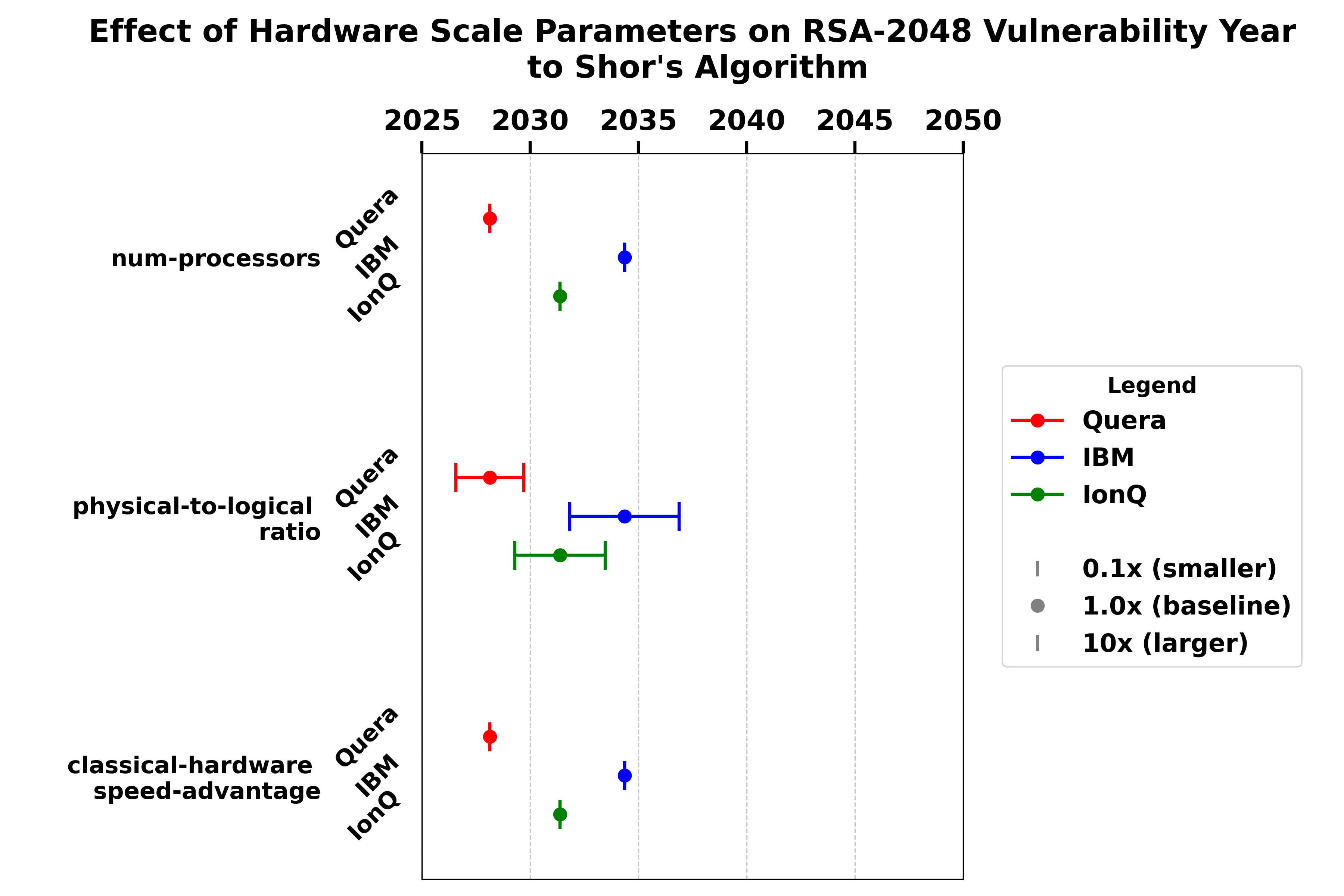}
\end{subfigure}
\caption{Robustness studies for Shor's algorithm using vendor roadmaps. The year of advantage (ie, physically feasible and economically advantageous) for Shor-2048 is plotted for each hardware roadmap and each input parameter. Bars show how the year of QEA changes when that parameter changes from the baseline. The chart on the left covers rates of improvement. The right chart shows the classical speed advantage, number of processors, and physical to logical qubit ratio changed by a factor of 10 smaller and larger. Note we find some roadmaps overly optimistic.}
\label{fig:shor_parameter}
\end{figure*}

\subsection{Predictions for Shor-2048 Are More-Robust}
In general, the predictions based on company roadmaps are somewhat optimistic (particularly for neutral atom-based hardware) but in line with previous predictions. Google researchers have mentioned the mid-2030s as a possible date \cite{hollister2024google}. However, Sevilla et al. in 2020 \cite{sevilla2020forecasting} found RSA-2048 unlikely to be solved before 2039. 

Our results for Shor-2048 are in general more robust to variations in our parameters (See Figure~\ref{fig:shor_parameter}). This is primarily due to the fact that advantage for Shor's algorithm is less limited by quantum overhead. In other words, once quantum computers with large numbers of qubits are developed they will be able to solve previously unsolvable factorization problems immediately. 
However, algorithms with less of an asymptotic advantages will not only need sufficient qubits but also incredibly fast quantum hardware at smaller problem sizes. If a problem does not immediately have a quantum advantage given its size or speed of quantum hardware, it will take many years to see an advantage for that problem size. For instance, given the slowdown in ion-trap systems it will take many years of quantum progress $10\%$ faster than classical progress to outcompete classical computers at search problems. This is a tall order given the incredible improvements in classical hardware (particularly parallel hardware such as GPUs) \cite{Rupp2018MicroprocessorTrendData}. These facts mean, that our analysis for Grover's algorithm is heavily dependent on the quantum overhead and the problem size considered. In addition, predictions that occur further out in our model are more dependent on initial conditions and improvement rates.

Conversely, for large-size problems and algorithms with larger asymptotic advantages (e.g, Shor), qubit-roadmaps become the primary constraint. This leads to the much larger prediction variation for parameters that affect the number of qubits, for instance, the physical to logical ratio (see \ref{fig:shor_parameter}).

\section{Limitations}
Our model has several key limitations. First, we suspect that many of the roadmaps we present in the tool may be somewhat over-optimistic. We invite others to use the tool with their own assumptions and compare with the default roadmaps. 

Second, we do not know the trend rates in classical hardware speed advantage (i.e, quantum slowdown) and classical compute cost advantage (i.e, cost overhead). However, our tool's key advantage is its ability for the user to enter in information that matches their priors. Further, many of our predictions are not sensitive to these parameters (see Section~\ref{Robustness Analysis}). 

We also must note that for an unstructured search on classical data, there are large data loading issues. Therefore, search problems over classical data necessitate QRAM \cite{Giovannetti_2008}.

Finally, we emphasize that QEA year, highlighted in the calculator  represent the first year any quantum advantage could be found. This does not represent the year at which an advantage for a given problem size can be found (for instance RSA-2048). Users can look at the graphs the tool generates to see when a problem relevant to their use case is efficient on a quantum computer.

\section{Conclusion}
The Quantum Economic Advantage Calculator introduced in this work provides a practical and interactive tool for individuals interested in exploring how different approaches to quantum development influence the timeline for transitioning to quantum hardware. Building on the principles of the Classical Hare and Quantum Tortoise framework, this tool transforms theoretical concepts into a tangible model that individuals can customize and analyze.

A key focus of the framework was enhancing user expressibility. Nearly all parameters affecting QEA are fully customizable, allowing users to tailor the model to reflect their assumptions and explore various scenarios. By incorporating several new inputs and capabilities not previously considered, the tool expands the scope of QEA analysis and provides a richer understanding of the factors driving quantum advantage.

Initial analysis points to important applications of quantum computing in the 2030s. Further, the tool brings specificity to the discussion around which factors matter most for building quantum computation. It depends on the algorithm: numbers of qubits matter most for some problems (namely Shor's algorithm) and quantum computer speed matters more for others (namely Grover's search algorithm).

Ultimately, the significance of this framework lies in its adaptability and accessibility. By offering users the freedom to perform their own analyses and draw their own conclusions, the tool empowers the quantum community, researchers, and the world  to better understand and navigate the evolving landscape of quantum computing.

\bibliographystyle{IEEEtran}
\bibliography{references}

\end{document}